\documentclass{aa}

\pdfoutput=1

\usepackage[varg]{txfonts}
\usepackage[utf8]{inputenc}
\usepackage{natbib}
\usepackage{graphicx}
\usepackage{amsmath}
\usepackage[dvipsnames]{xcolor}
\usepackage{siunitx}
\usepackage{float}
\usepackage[colorlinks]{hyperref}

\begin{document}

\title{A Bayesian estimation of the Milky Way's circular velocity curve using Gaia DR3}

\author{
Sven P\~oder\inst{1}\inst{2}
\and
{Mar\'ia Benito}\inst{3}\inst{1}
\and
{Joosep Pata}\inst{1}
\and
{Rain Kipper}\inst{3}
\and
{Heleri Ramler}\inst{3}
\and
{Gert H\"utsi}\inst{1}
\and
{Indrek Kolka}\inst{3}
\and
{Guillaume F. Thomas}\inst{4}\inst{5}
}

\institute{
NICPB, R\"avala 10, Tallinn 10143, Estonia \email{sven.poder@kbfi.ee}\label{inst1}
\and
Tallinn University of Technology, Ehitajate tee 5, Tallinn 19086, Estonia\label{inst2}
\and
Tartu Observatory, University of Tartu, Observatooriumi 1, T\~oravere 61602, Estonia \email{mariabenitocst@gmail.com} \label{inst3}
\and
Instituto de Astrofísica de Canarias, C/Vía Láctea s/n, 38205 La Laguna, Tenerife, Spain\label{inst4}
\and
Universidad de La Laguna, Dpto. Astrofísica, Avenida Astrofísico Francisco Sánchez, 38206 La Laguna, Tenerife, Spain\label{inst5}
}

\abstract{}
   {Our goal is to calculate the circular velocity curve of the Milky Way, along with corresponding uncertainties that quantify various sources of systematic uncertainty in a self-consistent manner.
   }
   {The observed rotational velocities are described as circular velocities minus the asymmetric drift. The latter is described by the radial axisymmetric Jeans equation. We thus reconstruct the circular velocity curve between Galactocentric distances from 5 kpc to 14 kpc using a Bayesian inference approach. The estimated error bars quantify uncertainties in the Sun's Galactocentric distance and the spatial-kinematic morphology of the tracer stars. As tracers, we used a sample of roughly 0.6 million stars on the red giant branch stars with six-dimensional phase-space coordinates from Gaia data release 3 (DR3). More than 99\%  of the sample is confined to a quarter of the stellar disc with mean radial, rotational, and vertical velocity dispersions of $(35\pm 18)\,\rm km/s$, $(25\pm 13)\,\rm km/s$, and $(19\pm 9)\,\rm km/s$, respectively.
   } 
   {We find a circular velocity curve with a slope of $0.4\pm 0.6\,\rm km/s/kpc$, which is consistent with a flat curve within the uncertainties. We further estimate a circular velocity at the Sun's position of $v_c(R_0)=233\pm7\, \rm km/s$ and that a region in the Sun's vicinity, characterised by a physical length scale of $\sim 1\,\rm kpc$, moves with a bulk motion of $V_{LSR} =7\pm 7\,\rm km/s$. Finally, we estimate that the dark matter (DM) mass within 14 kpc is $\log_{10}M_{\rm DM}(R<14\, {\rm kpc})/{\rm M_{\odot}}= \left(11.2^{+2.0}_{-2.3}\right)$ and the local spherically averaged DM density is $\rho_{\rm DM}(R_0)=\left(0.41^{+0.10}_{-0.09}\right)\,{\rm GeV/cm^3}=\left(0.011^{+0.003}_{-0.002}\right)\,{\rm M_\odot/pc^3}$. In addition, the effect of biased distance estimates on our results is assessed.  

   }
    {}


\maketitle

\section{Introduction}

The rotation of stars and gas in galactic discs has been extensively used as a kinematical tracer of matter distribution of external galaxies and our own Galaxy, the Milky Way (MW) (\citealt{2022arXiv220104136K}). 
Several recent studies (\citealt{2019ApJ...870L..10M, 2019ApJ...871..120E, 2020A&A...642A..95C, 2020ApJ...895L..12A, 2022arXiv220413672K, 2022arXiv220606207G, 2022arXiv221105668W, 2022arXiv221210393Z}) have measured the stellar disc rotation in the MW using stellar data from the Gaia satellite (\citealt{2016A&A...595A...1G}). These studies differ in the samples of stars used as a tracer and/or in the methodology and assumptions employed. Moreover, some of the cited studies provided rotational (azimuthal) velocities, whereas the others presented circular ones. 
The former are direct measurements and no underlying assumptions are made with respect to the shape or time dependence of the MW's gravitational potential. On the other hand, circular velocities assume a stationary gravitational potential that exhibits axial symmetry. Moreover, these are the velocities that should be used to derive the total or dynamical matter distribution. The modelling assumptions can therefore bias the determination of the total and dark matter content in our Galaxy. 

The amount of phase space data currently available is large, and statistics is generally not the limiting factor for studies of the dynamics of the MW stellar disc. The limiting factor is instead systematic errors, such as the Sun's Galactocentric distance or the adoption of incorrect modelling assumptions. In this paper, we present a Bayesian inference approach to estimate the circular velocity curve of our Galaxy that allows the straightforward incorporation of systematic and statistical sources of uncertainty, which are both treated as nuisance parameters. In this way, we provide, for the first time, a circular velocity curve with errorbars that self-consistently include uncertainties in the Sun's Galactocentric distance and in the spatial-kinematic structure of the stellar disc. Therefore, our uncertain knowledge about astrophysical parameters is propagated through Bayes' theorem to the determination of the circular velocity curve and, subsequently, to the estimation of the dark matter density profile in our Galaxy. Taking into account the uncertainties about how dark matter is distributed in the MW is essential for interpreting the results of dark matter particle searches (see e.g. \citealt{2017JCAP...02..007B}).

The Bayesian inference approach presented here is a flexible method that models the observed rotational or azimuthal velocity at a given Galactocentric distance as the difference between the circular velocity and the asymmetric drift component. The latter velocity component is obtained from the stationary, axisymmetric radial Jeans equation under the assumption of symmetry above and below the Galactic plane. Observed and modelled azimuthal velocities are then compared by means of the Bayes theorem. 
The paper is divided as follows. Section~\ref{sec:data} describes the data; Section~\ref{sec:method} presents the Bayesian methodology. The results are presented in~\ref{sec:circvel}, and we conclude in Section~\ref{sec:conclusions}.

\section{Data}
\label{sec:data}

\subsection{Red giant branch stars}
\label{subsec:RCsample}

We used astrometric data and radial velocities from Gaia data release 3 (DR3) for $665\,660$ stars in the red giant branch (RGB). 
These stars are old and have relatively large velocity dispersion. Thus, they are less susceptible to perturbations. They are also bright enough to have measured radial velocities out to large distances.
Specifically, we used the same sample of almost six million RGB stars as in~\cite{2022arXiv220606207G} to which we performed additional spatial and kinematic cuts.

In the following we briefly describe how the RGB sample was obtained and we refer readers to the original paper~\cite{2022arXiv220606207G} for a thorough description.
Red giants are identified as stars with effective temperatures between $3000\,{\rm K}$ and $5500\,\rm K$ and surface gravity satisfying the condition $\log g < 3$. Both stellar parameters are provided as data products in Gaia DR3~\citep{2022arXiv220606138A}. Using these first selection criteria, we obtained 11 576 957 sources. We then selected RGB stars with good astrometric data as measured by the fidelity parameter $f_a$ given in~\cite{2022MNRAS.510.2597R} and we removed stars with $f_a \leq 0.5$, thus remaining with 6 586 329 stars. After this, we performed a cut in height above and below the Galactic plane, $|z|<0.2\,\rm kpc$ (which removes ca. 4.5M stars), a cut in Galactocentric distances $\SI{5}{kpc} < R < \SI{14}{kpc}$ (removing ca. 84k stars), and a cut in heliocentric velocity $|\textbf{v}-\textbf{v}_{\odot}|<210\,\rm km/s$.\footnote{$\textbf{v}_{\odot}$ is defined in Eq.~\eqref{eq: v_odot}.} The latter cut removed roughly 28k stars. The $z$ and velocity cuts are applied to remove halo stars (\citealt{2018Natur.563...85H, 2019MNRAS.483.3119T}). 
The cut in height also avoids large density and velocity gradients in the z-coordinate, thus making the derivatives with respect to $z$ in the Jeans equation negligible (see Eq.~\eqref{eq:Jeans_eq_1}). We note that the scale height of the thin disc is roughly 250 pc~(\citealt{BlandHawthornGerhard16}).

\begin{figure*}[h]
    \centering\includegraphics[width=17cm]{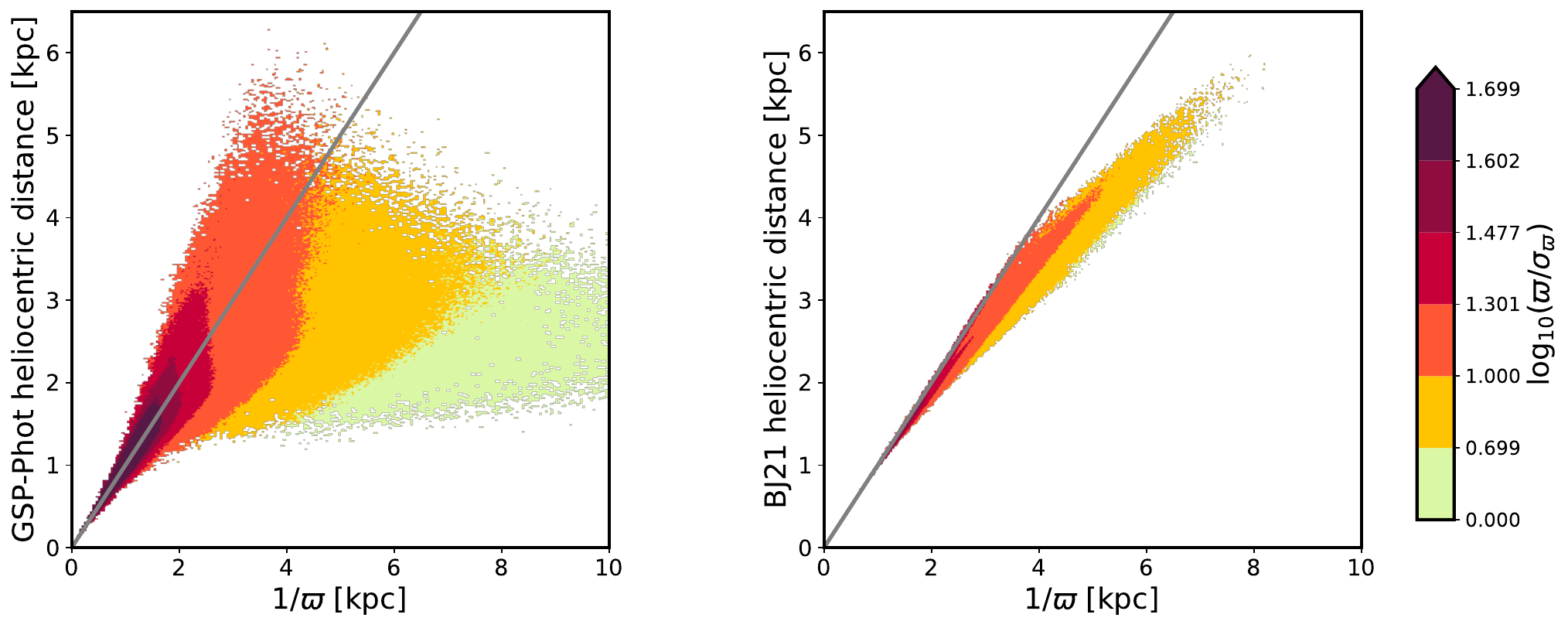}
    \caption{Bias in different distance estimates. Left: GSP-Phot heliocentric distances as a function of parallax inverse. The colour bar shows the mean parallax quality inside each pixel. The quality cut used in this paper corresponds to $\log_{10}\left(\varpi/\sigma_{\varpi}\right)>1.301$. Right: same but for photogeo heliocentric distances by~\cite{2021AJ....161..147BMr}.
    }
    \label{fig:heliocentric_distances}
\end{figure*}

\begin{figure*}[h]
    \centering\includegraphics[width=17cm]{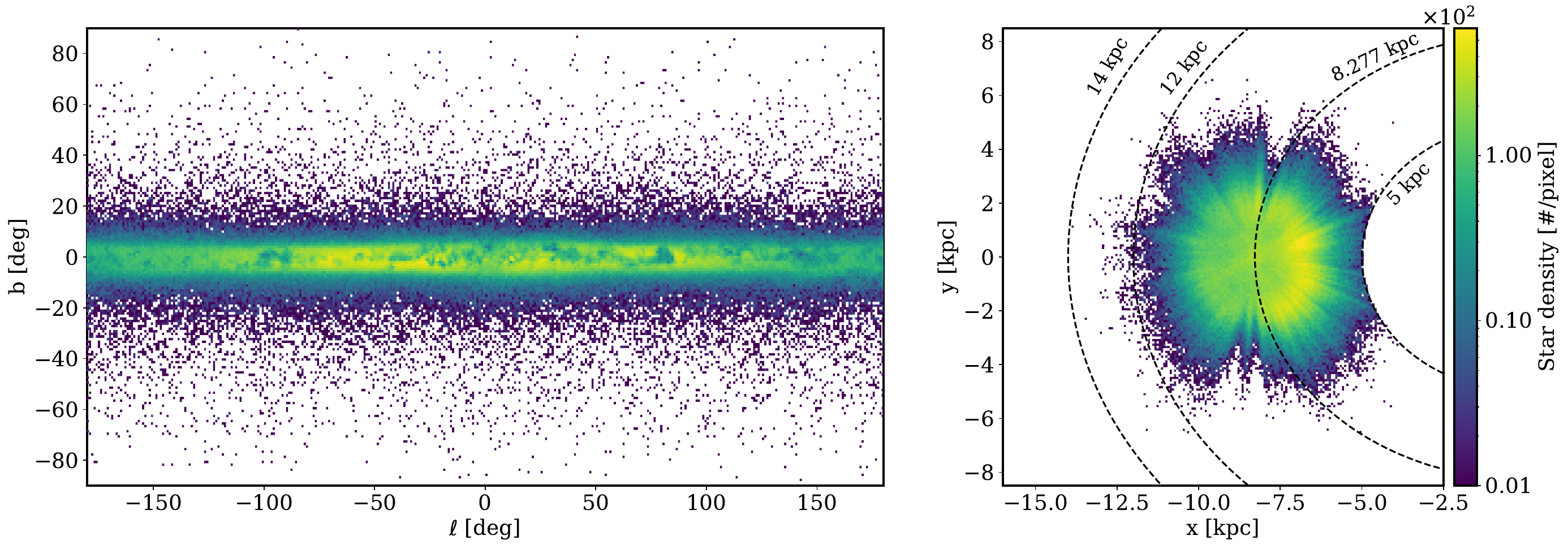}
    \caption{Spatial distribution of the final red giants sample. Left: distribution in the Galactic longitude ($\ell$) and latitude ($b$) plane. The colour bar shows the number density of stars per pixel. Each pixel has a size of 1 degree in both latitude and longitude. 
    Right: same but projected into the Galactic plane. Each pixel has a size of 0.05 kpc and 0.06 kpc in the x and y coordinates, respectively. In this figure, the Galactic centre is located at (0, 0), the Sun is located at $(-8.277, 0)$ and the rotation of the Galaxy is clockwise. We added dashed, black circles at 5 kpc, $R_0=\SI{8.277}{kpc}$, 12 kpc and 14 kpc to ease visualisation. 
    For the transformation to Galactocentric coordinades, we adopt $R_0=8.277\,\rm kpc$, $z_0=25\,\rm pc$ and ${\bf v}_\odot=(11.1, 251.5, 8.59)\, \rm km/s$.
    }
    \label{fig:RCs_lb}
\end{figure*}

We used GSP-Phot distances~\citep{2022arXiv220606138A} instead of the inverse of the parallax due to noisy parallax measurements. As shown in the left panel of Fig.~\ref{fig:heliocentric_distances}, the GSP-Phot distances are significantly underestimated at large distances from the Sun (see also \citealt{2022arXiv220606138A, 2022arXiv220605992F}). This would lead to artificially lower circular velocities and thus to a steeper slope of the estimated circular velocity curve. To reduce this bias, we imposed a tight constraint on the quality of the parallax measurements, namely $\varpi /\sigma_{\varpi } > 20$. This cut-off removed approximately 1.3M stars and suppressed the systematic underestimation of the distances, in fact leading to a slight overestimation. To assess the dependence of our results on inaccurate distances, we further performed the same analysis using a less stringent quality cut on parallaxes and  'photogeo' distances from~\cite{2021AJ....161..147BMr} (henceforth referred to as BJ21). Photogeo distances suffer from underestimation when including measurements with  $\varpi /\sigma_{\varpi } \lesssim 20$  (see the right panel of Fig.~\ref{fig:heliocentric_distances}). In the end, we are left with a final sample of $665\,660$ stars which is shown in Figure~\ref{fig:RCs_lb}. 

\begin{figure*}
    \centering\includegraphics[scale=0.25]{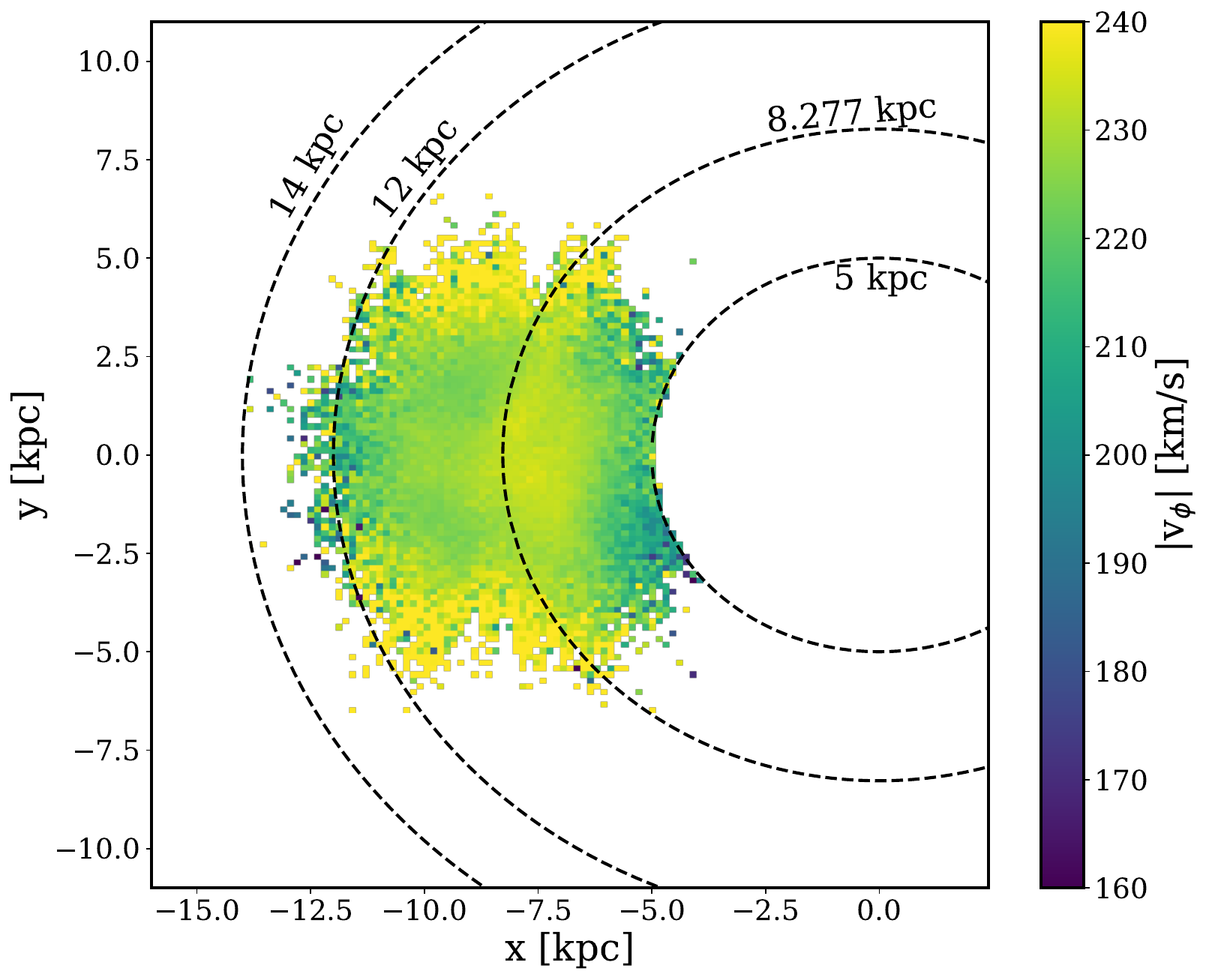} \hspace{1cm}
    \centering\includegraphics[scale=0.3]{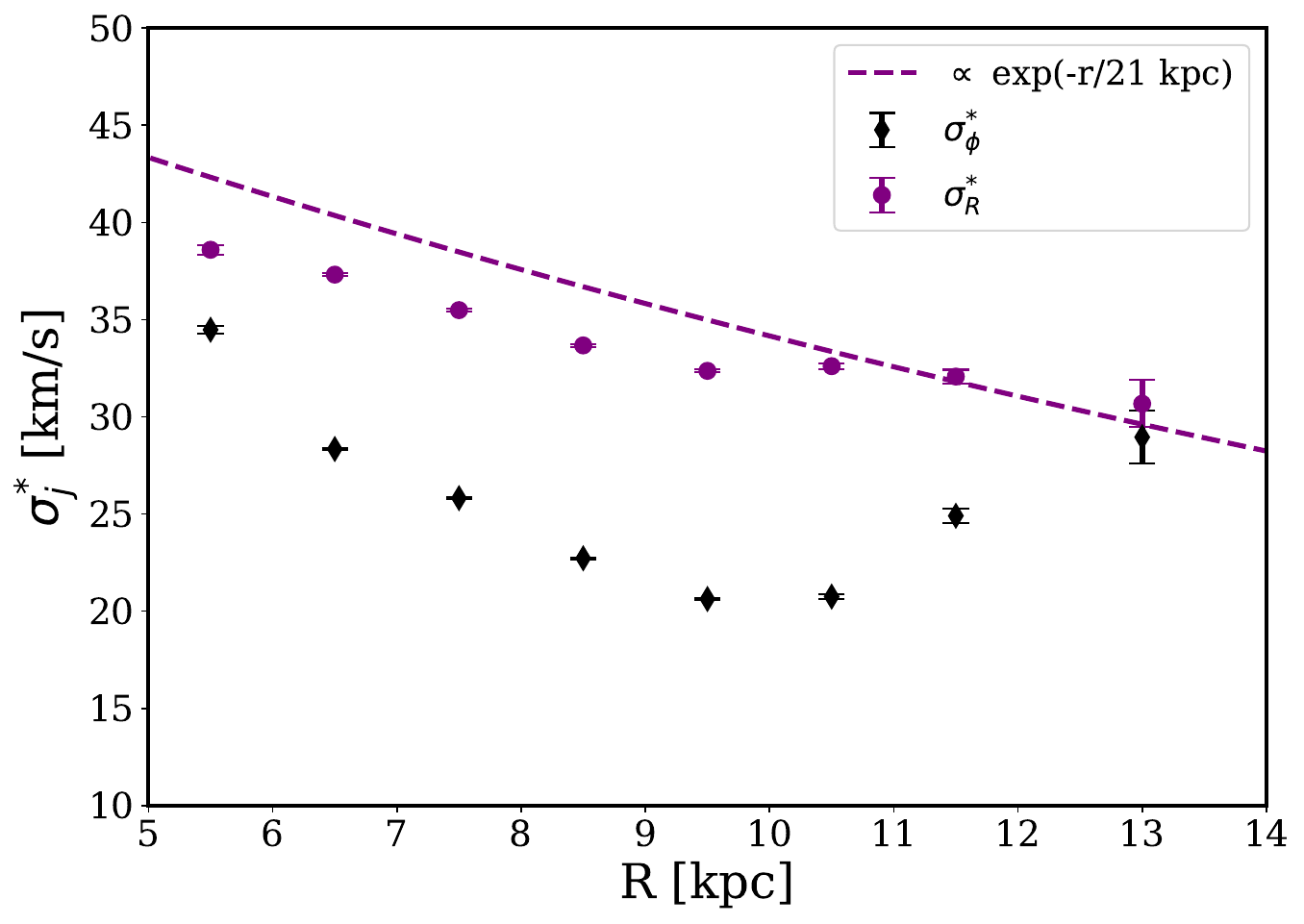}
    \caption{Distribution of the mean observed rotational velocity $v_\phi$ inside x-y pixels 
    for our final sample of RGB stars (left). The size of the bin is 150 pc in both x and y coordinates, respectively. We added dashed, black circles at 5 kpc, $R_0=\SI{8.277}{kpc}$, 12 and 14 kpc to ease visualisation. Square root of the radial and azimuthal diagonal components of the velocity dispersion tensor in radial bins of 1 kpc (right).
    }
    \label{fig:RCs_v_phi}
\end{figure*}

\subsection{Galactocentric frame}
\subsubsection{Galactocentric transformation}
\label{sec:transformation}
We transformed RGB stars from a heliocentric to a Galactocentric reference frame. 
We used a right-handed Galactocentric coordinate system $(x, y, z)$, with the Sun located at negative $x$, $y$ pointing in the direction of the Galactic rotation and $z$ towards the North Galactic Pole.
In order to define this new frame and perform the transformation correctly, we assumed solar orbital parameters ($R_0$, $z_0$, $U_\odot$, $V_\odot$, $W_\odot$) from contemporary literature. 
First, we treated the Galactocentric distance $R_0$ as a nuisance parameter of the analysis in order to account for uncertainties in its determination. In particular, we used a uniform prior range $R_0\in [7.8-8.5]\,\rm kpc$ that encompasses recent estimates with their corresponding uncertainties (\citealt{2019Sci...365..664D, refId0, Abuter:2018drb, 2019A&A...625L..10G, 2020A&A...636L...5G, 2022A&A...657L..12G, 2022MNRAS.tmp.3316L}). Our intention was not to constrain the value of $R_0$, but rather to show how the uncertainty in this parameter, which is encoded in the prior, propagates into the circular velocity curve. For this reason, we remained agnostic about the actual value of $R_0$ and adopted a uniform distribution that encompasses the most recent $R_0$ estimations within 2$\sigma$ uncertainty.
Second, for the height of the Sun over the Galactic plane $z_0$, we assumed a value of 25 pc (\citealt{2008ApJ...673..864J})\footnote{By adopting an alternative $Z_0$ value of 0 pc, the change in the central circular velocities is smaller than 1\%.}. 
The transformation from spherical coordinates in ICRS to a Galactocentric Cartesian setting was done as described in~\cite{1987AJ.....93..864J} and \cite{2018gdr2.reptE...3H}. 

Finally, the radial velocity measurements from Gaia also allow us to construct the full 3D space velocities and then transform them to a new frame by using the Sun's Galactocentric velocity.  
Under the assumption that Sg A$^*$ is at rest at the Galactic centre, the y and z components of the total solar velocity vector are derived from the proper motion of Sg A$^*$, as measured in~\cite{2020ApJ...892...39R}, in combination with the adopted value of $R_0$. On the other hand, for the x-component of the velocity, we adopted the value from~\cite{Schooenrich+2010}. 
We have not corrected this value for the offset in radial velocity between the radio-to-infrared reference frames determined by the GRAVITY collaboration (\citealt{2019A&A...625L..10G, 2020A&A...636L...5G, refId0, 2022A&A...657L..12G}) as suggested in~\cite{2018RNAAS...2..210D}. There are many possible sources for this systematic offset and, in any case, it is compatible with zero at the 2$\sigma$ level (\citealt{2022A&A...657L..12G}). 
In this way, we obtain the following vector
\begin{equation}
    \label{eq: v_odot}
    {\bf v}_\odot = \begin{bmatrix}U_\odot \\ V_\odot \\ W_\odot \end{bmatrix} = \begin{bmatrix}11.1 \\ 251.5 \times\left(\frac{R_0 }{\SI{8.277}{kpc}}\right) \\ 8.59 \times \left(\frac{R_0 }{\SI{8.277}{kpc}}\right)\end{bmatrix} \rm km/s,
\end{equation}

where $U_\odot$, $V_\odot$, $W_\odot$ correspond to the velocity components in the Galactocentric x, y, and z-directions, respectively. As $U_\odot$ corresponds to the radial motion of the Sun towards the Galactic centre, we implicitly assumed that the LSR has no such motion.

Using Gaia measurements for right ascension, declination, and radial velocity, and the GSP-Phot distances with a quality parallax cut of $\varpi/\sigma_{\varpi}>20$,
we transformed the proper motions and radial velocities first to Cartesian velocities in a similar way as was done in~\cite{1987AJ.....93..864J} and \cite{2018gdr2.reptE...3H}. Finally, we switched to Galactocentric cylindrical coordinates $(R, \phi, z, v_R, v_{\phi}, v_z)$.

\begin{figure*}
    \centering\includegraphics[scale=0.26]{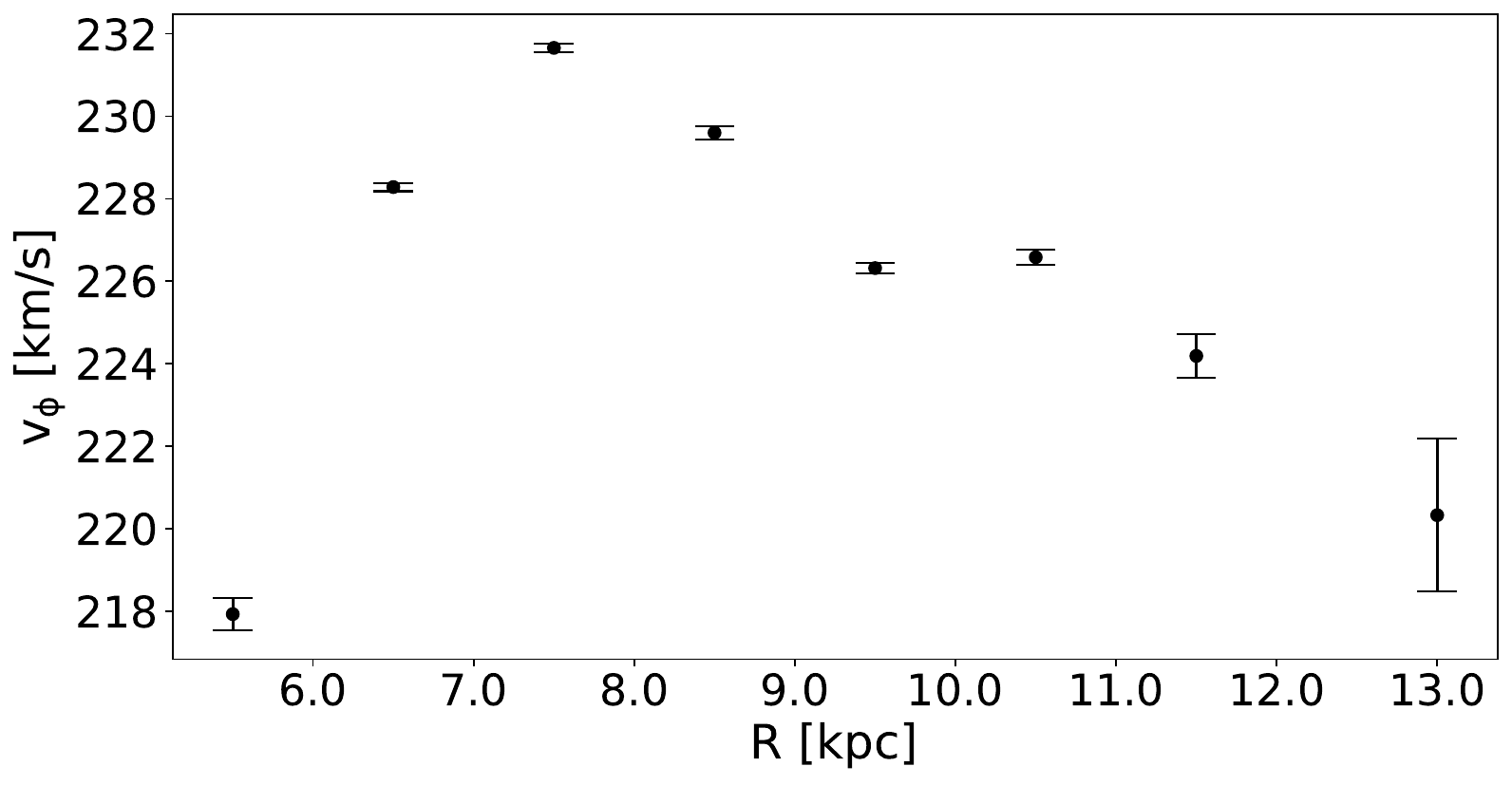} \hspace{1cm}
    \centering\includegraphics[scale=0.26]{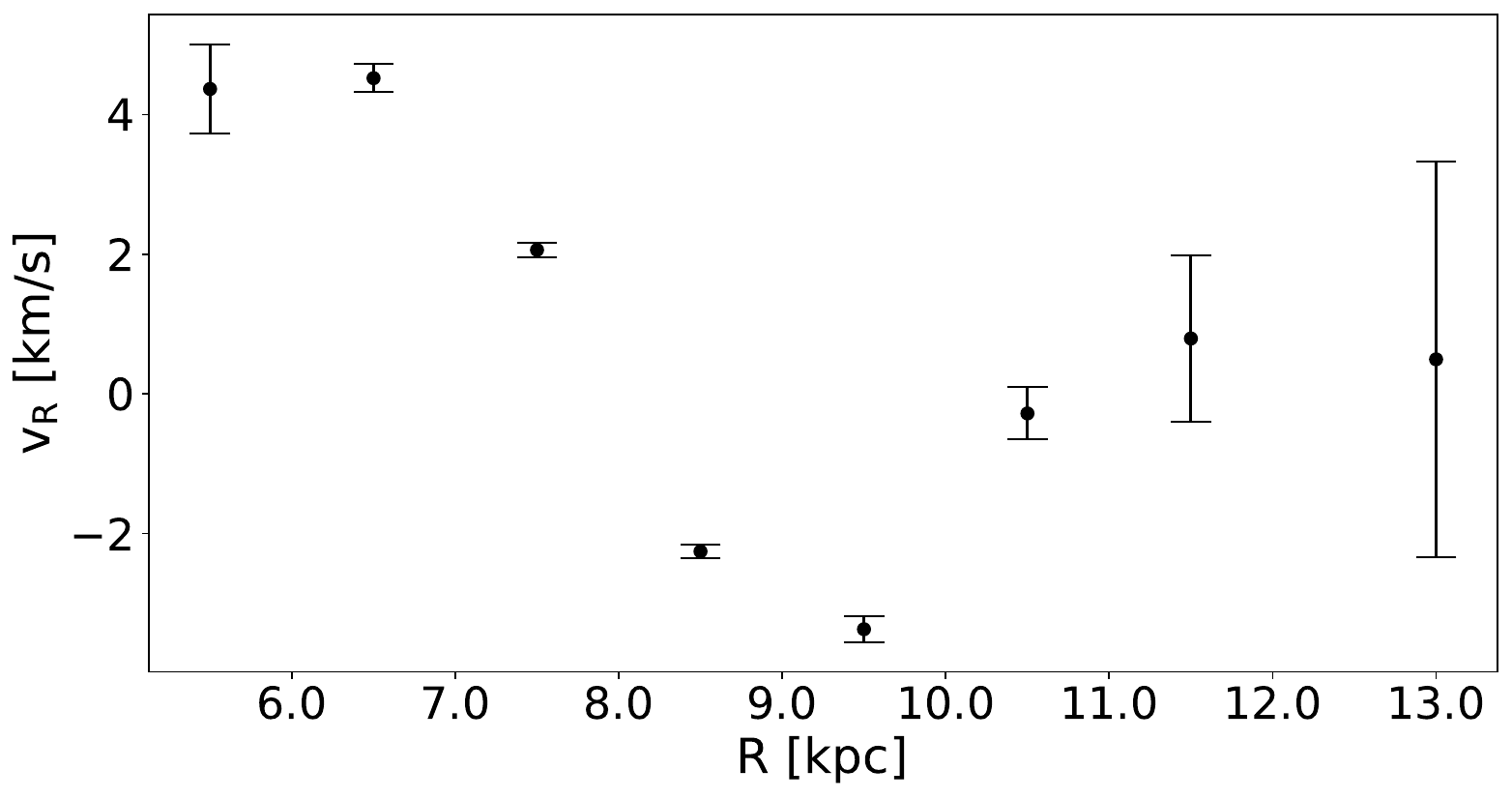}
    \caption{Mean azimuthal (left) and radial (right) velocities as a function of Galactocentric distance $R$. As in figure~\ref{fig:RCs_v_phi}, errorbars are calculated via bootstrap (see text for details). 
    }
    \label{fig:RCs_v_phi_r}
\end{figure*}

The left panel of figure~\ref{fig:RCs_v_phi} shows the mean observed rotational velocities in the Galactic plane. The right panel of the same figure depicts the mean axisymmetric radial and azimuthal velocity dispersions. Figure~\ref{fig:RCs_v_phi_r} shows the mean azimuthal and radial velocities. In both figures uncertainties are calculated by bootstrap resampling and are given by half the interval between the 16th and 84th percentiles of the corresponding distribution. As shown in the right panel of the last figure, the bulk motion of the stars in the radial direction exhibits an oscillatory pattern with an amplitude of roughly 5 km/s. This was already reported in Gaia DR2 \citep{2018A&A...616A..11G} and might be a kinematic signature of the spiral arms or the result of interaction with a perturber. We leave for future work a careful study of the origin of this intriguing oscillation in the radial velocities.

\subsubsection{Binning}
\label{subsubsection:binning}
We treated the Sun's Galactocentric distance as a free parameter in our analysis. Varying $R_0$ translates into a variation of the $R$ coordinate of the RGB stars. For this reason, we distributed the RGB sample in bins defined in the adimensional coordinate $x=R/R_0$. This mitigates the shift of the red giants' $R$ coordinate when varying $R_0$. Thus, for different $R_0$ values, a given $x$ bin contains approximately the same stars (\citealt{2019JCAP...03..033B}). 
Furthermore, we note that by marginalising over the azimuthal coordinate $\phi$, the rotational velocity in the Galactic disc is treated as a purely radially dependent observable. 

\begin{figure}
    \resizebox{\hsize}{!}{\includegraphics{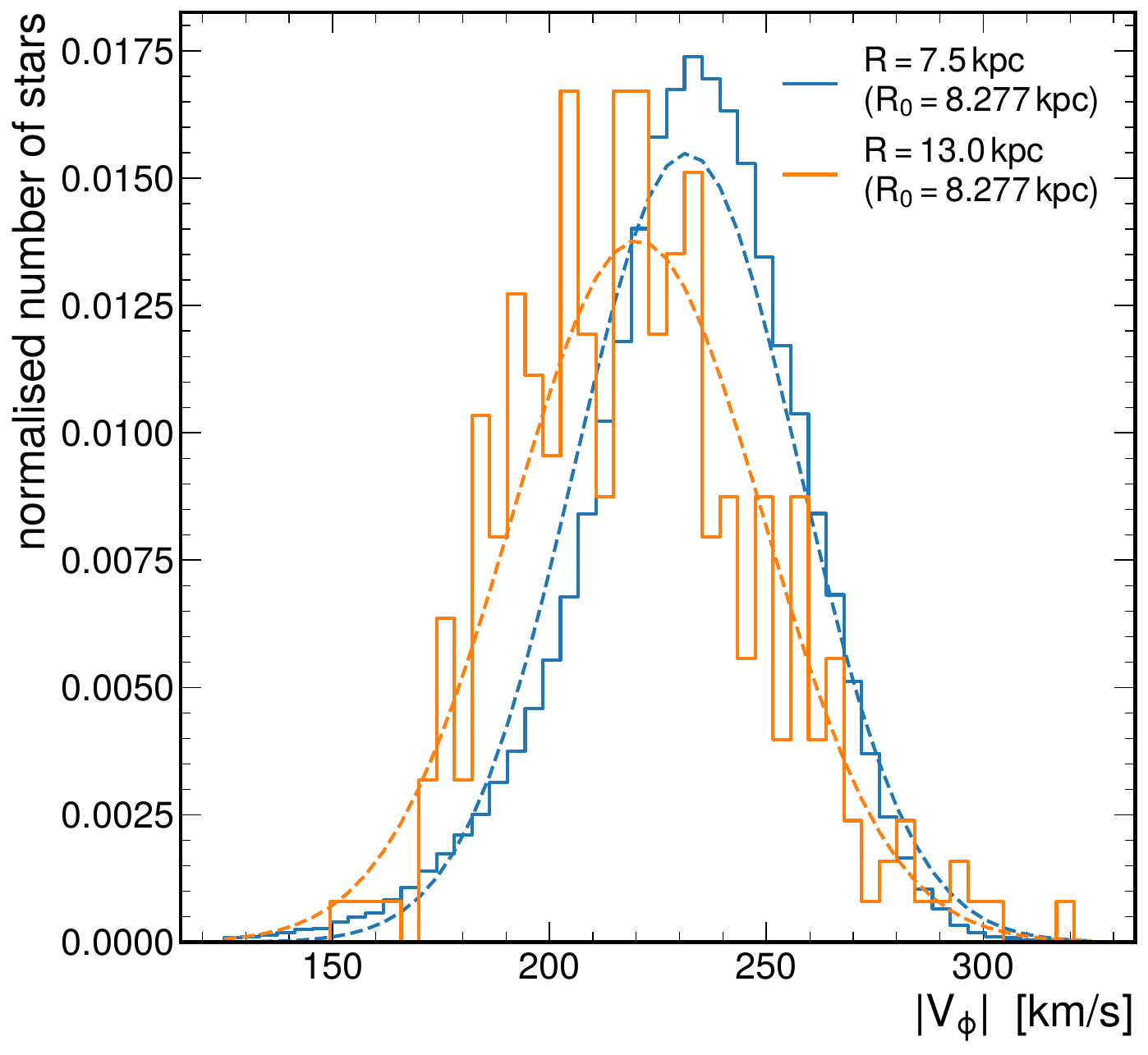}}
    \caption{Distribution of $v_\phi$ inside the last radial bin where the distribution of values deviates most from a Gaussian distribution (orange) and a randomly selected bin (blue). The dashed lines depict the best-fitted Gaussian of the rotational velocity inside each particular bin.}
    \label{fig:vphidist}
\end{figure}

In total, we defined eight bins from $x=5/8.277$ to $x=14/8.277$ with a step of $\Delta x=1/8.277$. Observed rotational (azimuthal) velocities inside each bin approximately follow a Gaussian distribution as shown in figure~\ref{fig:vphidist}. In this figure, we plot the distribution of observed velocities inside two bins: the bin where the distribution deviates most from a Gaussian and a randomly selected bin. The rotational velocity inside each bin is defined by the mean.
The associated uncertainties are calculated by bootstrap resamplings and are given by half the interval between the 16th and 84th percentiles of the velocity distributions.

\section{Methodology}\label{sec:method}

\subsection{Axisymmetric kinematic model}
\label{subsec:model}

\renewcommand{\thefootnote}{\roman{footnote}}

Inside each radial bin, we modeled the mean rotational or azimuthal velocity as 
\begin{equation}
\label{eq:model_azimuthal_velocity}
    v_{\phi{{\rm model}}} = v_c - v_a,
\end{equation}
where $v_c$ is the circular velocity or the velocity of a star moving in a circular orbit and $v_a$ is the asymmetric drift. The latter accounts for the diffusion of stars in phase-space as the stars orbit the Galaxy and streaming or bulk motions inside the disc. Neglecting large scale non-axisymmetric features, the asymmetric drift component can be obtained from the radial Jeans equation under the assumption that the MW is in a steady-state and has an axisymmetric gravitational potential (\citealt{2008gady.book.....B}). If we further expect the density distribution to be symmetric with respect to the Galactic plane, the Jeans equation takes the following form 
\begin{equation}\label{eq:Jeans_eq_1}
\frac{R}{\nu}\frac{\partial\left(\nu\overline{v_R^2}\right)}{\partial R} + R\frac{\partial \overline{v_Rv_z}}{\partial z} + \overline{v_R^2}-\overline{v_\phi^2}+v_c^2= 0.
\end{equation}
Substituting $\overline{v_\phi^2}=\sigma_\phi^{* 2}+(\overline{v_\phi})^{2}$~\footnote{To avoid confusion between uncertainties and the components of the velocity-dispersion tensor, we add the superscript $*$ to the latter as in~\cite{2022arXiv220606207G}.} and assuming $\overline{v_R^2}=\sigma_R^{*2}$, we further obtain

\begin{equation}
\label{eq:Jeans_traditional}
    \sigma_\phi^{*2} - \sigma_R^{*2}-\frac{R}{\nu}\frac{\partial\left(\nu\sigma_R^{*2}\right)}{\partial R}-R\frac{\partial \overline{v_Rv_z}}{\partial z} = v_c^2 - (\overline{v_\phi})^2.
\end{equation}
We have checked the latter assumption explicitly and observed that the measured values of $\overline{v_R^2}$ are at least two orders of magnitude larger than the measured $\overline{v_R}^2$. 

In traditional approaches, circular velocities in radial bins are calculated by plugging numbers into Eq.~\eqref{eq:Jeans_traditional}. Circular velocity values between bins are correlated as the radial dependence of the density profile and radial velocity dispersion are described by exponential functions. In the proposed new approach, we add Eq.~\eqref{eq:model_azimuthal_velocity} to transform the problem into an inference procedure. In this way, we can introduce free parameters such as scale lengths of the exponential functions and/or the Sun's Galactocentric distance $R_0$. The fitting procedure introduces an additional correlation between the bins. Nonetheless, we found that if we fix all free parameters of the analysis (i.e. scale lengths and $R_0$), the traditional and new approaches give the same circular velocity curve.

By expanding the difference of squares on the right hand side and introducing equation \eqref{eq:model_azimuthal_velocity}, we arrive to the following expression for the asymmetric drift

\begin{equation}\label{eq:full_asymmetric_drift_eq}
   v_a = \frac{\sigma_R^{*2}}{v_c + \overline{v_\phi}}\left[\frac{\sigma_\phi^{*2}}{\sigma_R^{*2}}-1-\frac{\partial \ln \nu}{\partial \ln R}-\frac{\partial \ln (\sigma_R^{*2})}{\partial \ln R}-\frac{R}{\sigma_R^{*2}}\frac{\partial \overline{v_Rv_z}}{\partial z}\right].
\end{equation}

The sum $v_c + \overline{v_\phi}$ in the denominator is often approximated as $\approx 2v_c$ (\citealt{2008gady.book.....B}). Nonetheless, we leave it as it is and estimate $\overline{v_\phi}$ as the mean rotational velocity inside each bin. In addition to this, the diagonal components of the velocity-dispersion tensor $\sigma_\phi^{*2}$ and $\sigma_R^{*2}$ are also calculated directly from the data, and they correspond to the variance of the azimuthal and radial velocity in the bin respectively. For the 3rd component inside the brackets of equation (\ref{eq:full_asymmetric_drift_eq}), the number density distribution $\nu$ is described by an exponential profile, namely $\nu\propto \exp{(-R/h_R)}$ with $h_R$ the disc scale radius. Notice that in the 4th component we describe $\sigma_R^{*}$ as $\sigma_R^{*}\propto\exp (-R/h_\sigma)$, where  $h_\sigma$ is the scale length of the radial velocity dispersion. Finally, after taking the derivatives with respect to $\ln R$ in (\ref{eq:full_asymmetric_drift_eq}), we are left with the following equation

\begin{equation}
\label{eq:asymmetric_drift_w_scaleparams}
   v_a = \frac{\sigma_R^{*2}}{v_c + \overline{v_\phi}}\left[\frac{\sigma_\phi^{*2}}{\sigma_R^{*2}}-1+R\left(\frac{1}{h_r} + \frac{2}{h_\sigma}\right)\right].
\end{equation}

We neglect the last term of \eqref{eq:full_asymmetric_drift_eq} in our axisymmetric treatment of the rotation curve derivation as $\overline{v_Rv_z} \approx 0$. This is motivated because the radial and vertical motions are expected to decouple for circular orbits near the disc when the velocity ellipsoid is aligned with the Galactic plane (\citealt{bovybook}). In reality, however, this is not necessarily true (e.g. some general models are provided by \citealt{TempelTenjes, Kipper:2016}). In any case, our cut in z-coordinate $|z|<0.2\, \rm kpc$ minimises gradients in the vertical direction. Furthermore, the inclusion of this term changes the final circular velocity at the percent level, as shown in~\cite{2019ApJ...871..120E}.

\subsection{Circular velocity fitting}

We used the axisymmetric kinematic model described in the previous section to derive the circular velocity $v_{c,j}$ in each radial bin $j$. We approached this as a Bayesian inference problem, wherein we used a Markov chain Monte Carlo (MCMC) algorithm to sample the posterior probability of our model parameters $\theta$, namely the circular velocities $\{v_{c, j}\}$, $h_R$, $h_\Theta$ and $R_0$. 
According to Bayes' theorem, the posterior distribution of a set of model parameters $\theta$ given a particular set of data $D$ can be defined as 
\begin{equation}
\label{eq:bayesinf}
    p(\theta\vert{D}) = \frac{p({D}\vert\theta)p(\theta)}{p(D)},
\end{equation}
where $p(\theta)$ is the prior distribution function that contains a priori knowledge about the parameters, $p(D)$ is the Bayesian evidence which is an irrelevant normalisation constant in this context. Moreover, the likelihood function takes the form
\begin{equation}
    p(D|\theta) = -\prod_{j} \left[\frac{1}{\sqrt{2\pi\sigma_{\overline{v_{\phi,j}}}^2}} \exp\left( \frac{\left(\overline{v_{\phi,j}} - v_{\phi{{\rm model},j}}(\theta)\right)^2}{\sigma_{\overline{v_{\phi,j}}}^2}\right) \right],
\end{equation}
where $j$ is iterated over $R$ bins. We note that this equation assumes that rotational and azimuthal velocities in each of the bins are independent of each other. 
The terms $\overline{v_\phi}_{,j}$ and $\sigma_{\overline{v_\phi}_{,j}}^2$ are obtained from the data and are the mean and variance of the azimuthal velocity in the j-th bin respectively. 

For the prior distribution in \eqref{eq:bayesinf}, we defined flat priors, where the circular velocities are allowed within a range of $[-400, 400]$ km/s. In addition to the velocities, we defined naive priors for the scale length parameters $h_R=3\pm 1$ kpc and $h_\sigma=21\pm 1$ so to encompass values in the literature (\citealt{2019ApJ...871..120E, BlandHawthornGerhard16}). As mentioned previously, the Galactocentric distance $R_0$ was also treated as a free parameter of the analysis and was given a uniform prior within $[7.8, 8.5] \, \rm kpc$. Since the solar Galactocentric velocities $V_\odot$ and $W_\odot$ are scaled with $R_0$ (see Eq.~\eqref{eq: v_odot}), the chosen prior is reflected both in the median value and error bar of the circular velocity in a given bin.

Having defined our likelihood and prior functions to use in the fitting, we set up our MCMC algorithm using the python package \texttt{emcee} (\citealt{Foreman-Mackey_2013}). The parameter space of our model was explored using 48 independent walkers. All in all, we used 13 parameters in the fitting, where the first ten were circular velocities $v_{c,j}$ of the radial bins and the rest were the Sun's Galactocentric distance and the scale length terms. 

By treating the Sun's Galactocentric distance as a free parameter of the analysis we were required to repeat the coordinate and velocity transformation at each step in the MCMC. In addition to this, we also had to propagate the covariance information of each star resulting in each step of the MCMC being computationally expensive. In order to bring down the iteration time, we used \texttt{numpy} (\citealt{harris2020array}) and \texttt{cupy} (\citealt{cupy_learningsys2017}), which make it possible to implement the calculations on both CPU and GPU in an efficient vectorised form.

The use of GPUs was particularly well motivated, since the parameter and uncertainty propagation routines in our code consist largely of matrix operations with relatively large arrays. GPU-accelerated computing libraries (such as \texttt{cupy}) take advantage of the fact that modern GPUs have significantly more threads than a CPU and are thus better at parallelising certain computation routines than their CPU-counterparts. In the end, both \texttt{numpy} and \texttt{cupy} were utilised simultaneously and the MCMC routine easily parallelised across the available CPUs and GPU devices where the most computationally demanding aspects of the pipeline were handled by the latter. The full data was analysed by using two CPU cores per GPU and with a total of six GPUs the computation time for each step was brought down to $\approx 11 \, s$. This translates into a 6-fold speed increase when compared to running the code with just a single GPU and a 164-fold increase when running solely on CPUs. Using a single GPU for the full dataset described in this work, quickly leads to either out of memory issues or extremely long runtimes and thus it must be noted that the feasibility of the analysis was heavily dependent on the availability of multiple GPU devices and CPU cores. Our RGB sample of roughly 0.6 million RGB stars and the code used in our analysis can be found \href{https://zenodo.org/record/8014011}{in zenodo} and \href{https://github.com/HEP-KBFI/gaia-tools}{https://github.com/HEP-KBFI/gaia-tools}, respectively.

\section{Results}
\label{sec:circvel}

\subsection{Circular velocity curve}
\begin{figure*}[h]
    \centering\includegraphics[width=17cm]{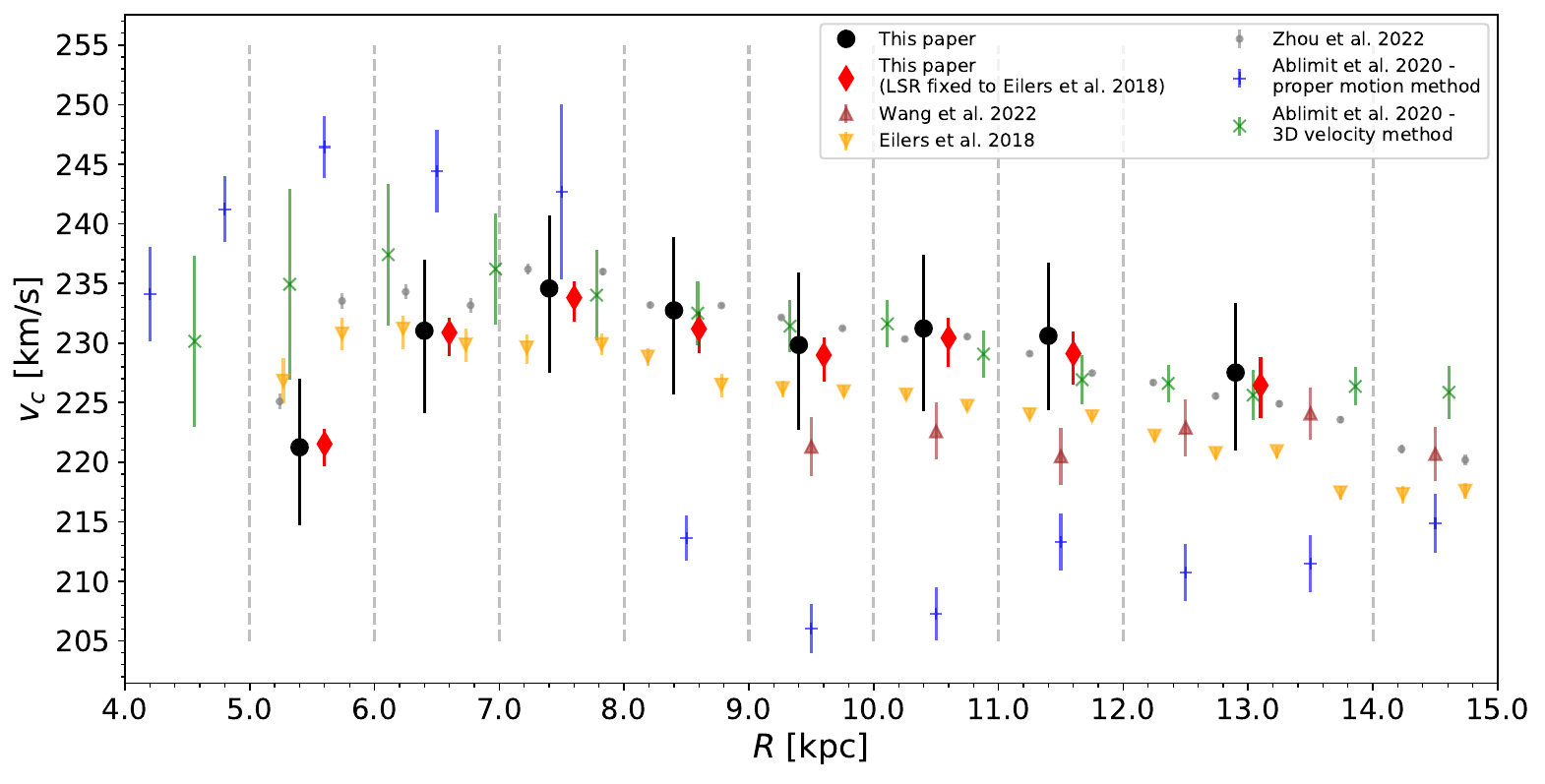}
    \caption{Circular velocity curve obtained from the MCMC. Grey dashed lines have been plotted to indicate the position of each radial bin. In black (with circles) we show the circular velocities as obtained in this paper where the error bars correspond to the 16th and 84th percentile of the circular velocity posterior distribution in a particular bin. We adopted $R_0=8.277\,\rm kpc$ to convert the adimensional coordinate $x$ into Galactocentric distance $R$.
    }    \label{fig:circvel}
\end{figure*}

The circular velocity curve of our sample of RGB stars is summarised in table~\ref{tab:circvel} and shown in figure~\ref{fig:circvel}. Inside each bin, we quote the median of the 1D marginalised posterior probability distribution obtained in the MCMC fitting. For the error bars, we quote the 16th and 84th percentile of the distribution. We would like to highlight that, in the classical approach for calculating the circular velocity curve, circular velocities in radial bins are calculated by plugging values into equation~\eqref{eq:Jeans_traditional}. In the proposed new approach, we add a simple kinematic model (given by \eqref{eq:model_azimuthal_velocity}) on top of the Jeans equation, thus transforming the problem into an inference procedure. This allows to introduce nuisance parameters, such as the $h_R$, $h_\sigma$ and $R_0$, and propagate their corresponding uncertainties (regardless of whether we have normal or non-normal errors) into the final circular velocity curve via Bayes theorem. The fitting procedure may, nonetheless, introduce additional correlations between the radial bins. For this reason, we checked that, if we fix the nuisance parameters $R_0$, $h_R$ and $h_\sigma$, the central values of the circular velocities obtained with the new approach (MCMC analysis) coincide with the values obtained by the classical or traditional technique.

In figure~\ref{fig:circvel}, we compare our result to others from the literature. 
Our circular velocity curve is in agreement with the one estimated in~\cite{2020ApJ...895L..12A} using the  3D velocity vector method on $\sim 10^3$ classical Cepheids. However, for $R>8\,\rm kpc$, we obtain larger circular velocities than those calculated by the same authors but using the proper motions of $\sim 600$ classical Cepheids. The former and latter samples have around 370 Cepheids in common and both results show that modelling assumptions and/or tracer samples can induce differences in the estimated circular velocities of at least 10\%. We note that these changes are larger than the estimated uncertainties in this work, which are in the $\lesssim 3$\% level. Our error bars include statistical uncertainties, which are negligible owing to the large data sample. They further include uncertainties in the spatial-kinematic morphology of the tracer stars (scale radius of the density profile $h_R$ and of the velocity distribution $h_\sigma$) and in the Sun's galactocentric distance. Circular velocities show a mild sensitivity to $h_R$, specially for values $h_R\leq 2.5\,\rm kpc$ and, at least within the prior range explored in our analysis, $h_\sigma$ and circular velocity central values are independent. On the contrary, the adopted value of $R_0$ strongly affects the final circular velocities and it is the main source of systematic uncertainties (from the ones studied in this analysis).

In addition, our estimated circular velocities are also compatible with those obtained in~\cite{2019ApJ...871..120E}, \cite{2022arXiv221105668W} and~\cite{2022arXiv221210393Z}, due to our large uncertainties compared to those estimated in these articles. If we fixed the Sun's galactocentric distance and total velocity in the azimuthal direction to the values adopted in the former article, namely $R_0=8.122\,\rm kpc$ and $V_\odot=245.8\,\rm km/s$, the estimated error bars on the circular velocities are reduced and our results are incompatible with \cite{2019ApJ...871..120E} analysis at 1$\sigma$ for our fiducial distance estimates (see section~\ref{results_distances} for a comparison using the circular velocity curve obtained with other distance estimates). This shows that $R_0$ is the main source of uncertainty in the reconstruction of the circular velocity curve.
Moreover, if for the fixed $R_0$ case the prior range in the scale length $h_\sigma$ is increased by a factor of three, the results remain unchanged. In contrast, increasing the prior range in the scale length $h_R$ by the same factor, the median values decrease by less than 2\% and the size of error bars remains roughly the same.

\begin{figure}[h]
    \resizebox{\hsize}{!}{\includegraphics{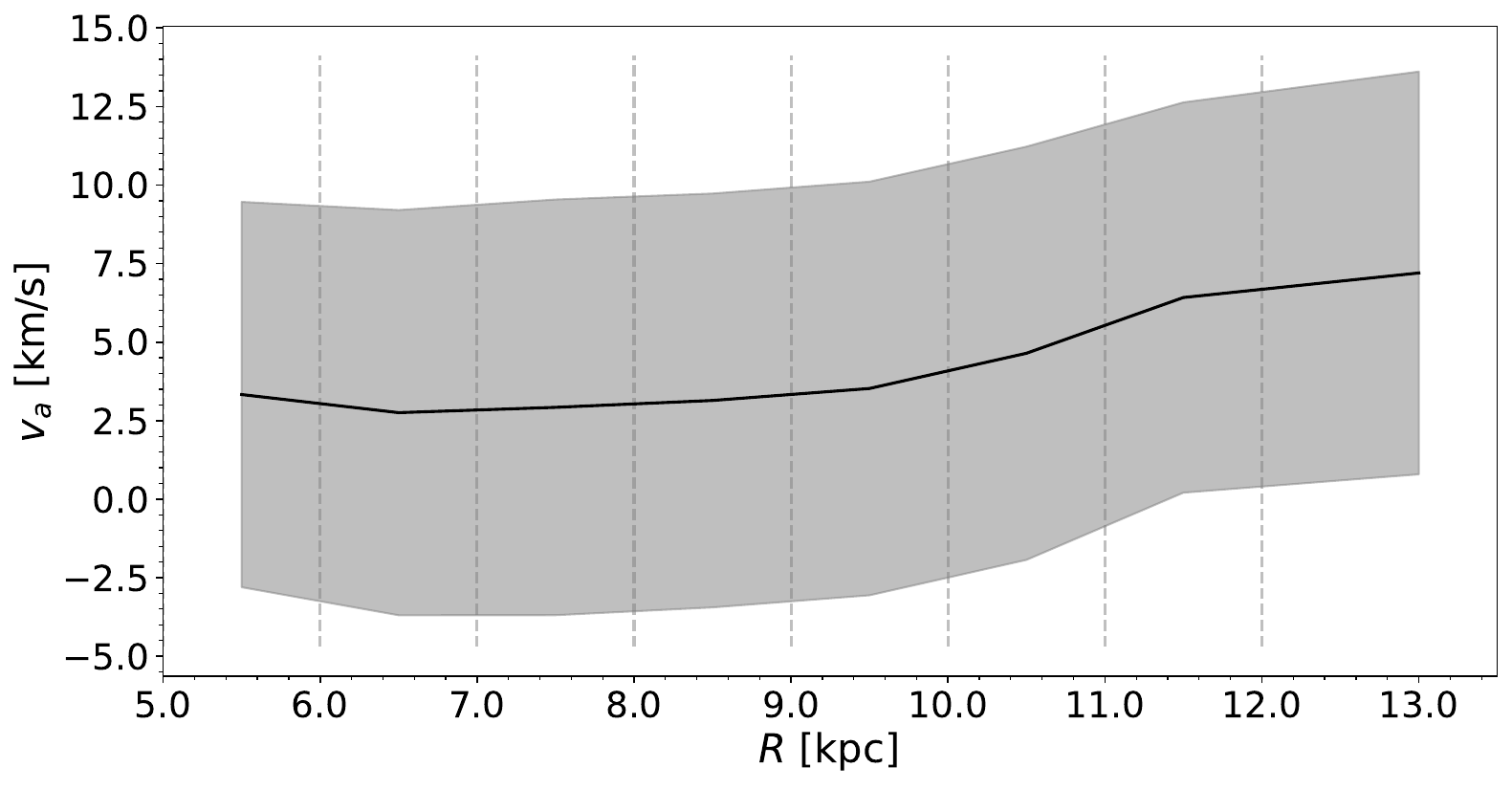}}
    \caption{Asymmetric drift profile. The black line shows the asymmetric drift correction in each radial bin and the shaded region depicts its propagated uncertainty. We note that the uncertainties are fully correlated between the bins.}
    \label{fig:va_r_plot}
\end{figure}

The circular velocity curve in~\cite{2022arXiv221105668W} was obtained by describing, by means of the radial axisymmetric Jeans equation, the dynamics of all Gaia DR3 stars within the region $160^{\rm o}<\ell < 200^{\rm o}$ and $|Z|<3\,\rm kpc$ that have measured radial velocities. All stars are thus described by the same asymmetric drift. However, younger stars are expected to have a smaller asymmetric drift than an older population of stars. In fact,~\cite{2019MNRAS.482...40K} estimated $v_a(R_0)=0.28\pm0.20\,\rm km/s$ using young classical Cepheids, whether we obtain, as expected, the central larger value $v_a(R_0)=3 \pm 7 \rm \,km/s$ for older RGB stars. Figure~\ref{fig:va_r_plot} shows the asymmetric drift as a function of Galactocentric distance for $R_0=8.277 \, \rm kpc$. The asymmetric drift mildly increases with distance to the Galactic centre with a slope of $0.59 \pm 0.12\,\rm km/s/kpc$.

Our estimated value of the circular velocity at the Sun's position, namely $233\pm 7\,\rm km/s$, is compared in table~\ref{tab:circvel_atSun} with other estimates from the literature.
We found that the estimated gradient of the curve is extremely sensitive to the radial interval included in its inference. If we remove the first two radial bins where the circular velocities increase, the obtained value is $-1.1 \pm 0.3 \, \rm km s^{-1} kpc^{-1}$, which points to a smooth decrease of the circular velocities with Galactocentric distances. If we rather include all radial bins, the estimated value of the slope is $0.4 \pm 0.6 \, \rm km s^{-1} kpc^{-1}$, which describes a flat circular velocity curve within the uncertainties. \cite{2019ApJ...870L..10M} and \cite{2019ApJ...871..120E} included all radial intervals for the determination of the slope, and found decreasing slopes that do not agree with the latter value. This may point out to the presence of systematic biases in the distance estimations, as described in section~\ref{results_distances}.  

\begin{table*}
    \caption{Measured circular velocities $v_c$ , also plotted in figure~\ref{fig:circvel}. We quote the median of each $x$ bin, as the fitting is done using this adimensional variable, and the corresponding $R$ value is calculated for $R_0=8.277\,\rm kpc$ (\citealt{2022A&A...657L..12G}).}
    \centering
    \begin{tabular}{c c c c c c c}
    \hline\hline
    x & R [kpc]& $v_c$ [km/s] & $\sigma_v^-$ [km/s]& $\sigma_v^+$ [km/s] & $v_a$ [km/s] & $\sigma_{va}$ [km/s] \\
    \hline
   0.66 & 5.5 & 221.3 & 6.5 & 5.7 & 3.3 & 6.1\\
   0.79 & 6.5 & 231.0 & 6.9 & 6.0 & 2.8 & 6.5\\
   0.91 & 7.5 & 234.6 & 7.1 & 6.1 & 2.9 & 6.6\\
   1.03 & 8.5 & 232.7 & 7.1 & 6.1 & 3.1 & 6.6\\
   1.15 & 9.5 & 229.8 & 7.1 & 6.1 & 3.5 & 6.6 \\
   1.27 & 10.5 & 231.2 & 7.0 & 6.2 & 4.6 & 6.6\\
   1.39 & 11.5 & 230.6 & 6.3 & 6.1 & 6.4 & 6.2\\
   1.57 & 13.0 & 227.5 & 6.5 & 5.8 & 7.2 & 6.4\\
    \hline
    \end{tabular}
\end{table*}\label{tab:circvel}

\subsection{Smooth dark matter halo}
In the region analysed in this article, mean radial velocities are $<5\%$ of the azimuthal or rotational velocities. According to~\cite{2020A&A...642A..95C}, when the ratio of radial to azimuthal velocities is smaller than 10\%, circular velocities obtained through the radial axisymmetric and stationary Jeans equation provide an unbiased estimator for the centrifugal velocities that balance the averaged radial gravitational force. According to this result, our circular velocity curve then provides an unbiased estimation of the spherically averaged dynamical mass distribution within $14\,\rm kpc$. Although this conclusion will be tested in an upcoming paper where we assess the effects of modelling assumptions, such as the axial symmetry condition, and the effect of Galactic substructure, using our current results we estimate that the DM mass is 
\begin{equation}
\log_{10}\left[M_{\rm DM}(R<14\, {\rm kpc})/{\rm M_{\odot}}\right]= 11.2^{+2.0}_{-2.3}.
\end{equation}
Furthermore, we find that the local spherically averaged DM density is 
\begin{equation}
\rho_{\rm DM}(R_0)=\left(0.41^{+0.10}_{-0.09}\right)\,{\rm GeV/cm^3}=\left(0.011^{+0.003}_{-0.002}\right)\,{\rm M_\odot/pc^3}.
\end{equation}
These estimates were obtained by fitting the observed circular velocities to the velocities predicted by Newtonian gravity for the baryonic components (stellar bulge, disc and gas) and the DM halo. For each baryonic component we adopted a set of three-dimensional density profiles, originally compiled in~\cite{2015NatPh..11..245I}. The stellar bulge mass is constrained by microlensing towards the Galactic centre and the stellar disc is normalised by the stellar surface density at the Sun's position. We describe the DM distribution using a generalised Navarro-Frenk-White density profile~\cite{1996MNRAS.278..488Z}. We compare observed and predicted velocities using the Bayesian prescription presented in~\cite{2019JCAP...09..046K, 2020JCAP...05..033K}. The estimates provided are Bayesian model averages that include uncertainties in the Sun's Galactocentric distance, the three-dimensional density profile of bulge and disc stars, and the stellar mass of the Galaxy.

We would like to highlight that our estimate of the local DM density is compatible, within $1\sigma$ uncertainties, with recent estimates of this quantity using the circular velocity curve method~\citep{2019ApJ...871..120E, 2019ApJ...870L..10M, 2019JCAP...10..037D, 2019MNRAS.487.5679L, 2019JCAP...09..046K, 2020Galax...8...37S}. In addition, it is compatible with local estimates using the vertical Jeans equation~\citep{2020A&A...643A..75S, 2020MNRAS.495.4828G, 2020MNRAS.494.6001N} and with the most-recent local estimate using a novel machine learning approach by~\cite{2023arXiv230513358L}. Thus reinforcing the conclusion about the spherical shape of the inner $\sim~15$ kpc of the DM halo, obtained by modelling stellar streams~\citep{2010ApJ...712..260K, 2015MNRAS.449.1391B} and the kinematics of halo stars~\cite{2019MNRAS.485.3296W}.

\subsection{The local standard of rest and the solar peculiar velocity }

The total Galactocentric azimuthal velocity of the Sun can be used to derive the solar peculiar velocity when we incorporate knowledge about the circular velocity at its position. In particular, the total azimuthal velocity is often decomposed as
\begin{equation}
\label{eq:solarvelocity}
    V_\odot = v_c (R_0)+ V_{\odot, LSR},
\end{equation}
where the last term is the Sun's peculiar motion in the local standard of rest (LSR).
The treatment of the solar velocity as shown in equation~\eqref{eq:solarvelocity} assumes that the LSR moves in a circular orbit about the Galactic centre and therefore, it coincides with the rotational standard of rest (RSR) in which stars move on circular orbits in the azimuthally averaged gravitational potential. However, in recent years it has been shown that the stellar disc exhibits bulk motions at the kiloparsec scale (\citealt{Bovy_2015_streaming_motions, Williams_streaming_motions, 2022arXiv220413672K}).
In the presence of these large scale streaming motions, the LSR, which is defined as the reference frame of a local population of stars with zero velocity dispersion, might not coincide with the RSR.
Considering this, the total azimuthal velocity of the Sun can be decomposed as (\citealt{2018RNAAS...2..210D})
\begin{equation}
    V_\odot = v_c(R_0) + V_{LSR} + V_{\odot, LSR},
\end{equation}
where $V_{LSR}$ is the velocity of the LSR with respect to the RSR. This difference of velocity between the LSR and RSR might account for the discrepancy between locally-derived estimations of the Sun's peculiar motion (i.e. using the Strömberg relation) and globally-measured values using a sample of tracers in a larger volume around the Sun. In fact, \cite{2012ApJ...759..131B} concluded that the LSR itself might not be on a circular orbit and it is rotating  $V_{LSR}\approx 12$ km/s faster than the actual RSR. This is in agreement with the recently reported value in~\cite{2022arXiv220413672K} of $\approx 10\,\rm km/s$. On the other hand, \cite{BlandHawthornGerhard16} estimated $V_{LSR}=0\pm\SI{15} {km/s}$. 

\renewcommand{\arraystretch}{1.4}
\begin{table*}
    \centering
    \caption{Circular velocity at the solar location $v_c(R_0)$ as measured by different methods. In the last column, we quote the value of the Sun's galactocentric distance that was adopted in each of the referenced articles. }  
    \begin{tabular}{c c c}
    \hline\hline
    Source & $v_c(R_0)$ [km/s] & $R_0$ [kpc] \\
    \hline
    This work & $233\pm 7$ & 8.277 \\
    
    \cite{2022arXiv221210393Z} & $234.04 \pm 0.08{\rm (stat.)} \pm 1.36{\rm (sys.)}$ & $8.122 \pm 0.031$ \\
    
    \cite{Kipper_nonstat} &  $228.4\pm3.5$ & 8.3 \\
    
    \cite{2019ApJ...871..120E} & $229.0 \pm 0.2$ & $8.122 \pm 0.031$ \\
    \cite{10.1093/mnras/sty2623} & $236 \pm 3$ & $8.2 \pm 0.1$\\
    
    \cite{Bobylev2017-zx} & $231 \pm 6$ & 8 \\
    \cite{10.1093/mnras/stw2096}& $240 \pm 6$ & 8.34 \\
    
    \cite{2012ApJ...759..131B} &  $218 \pm 6$ & $8.1_{-0.1}^{+1.2}$ \\
   
    \hline
    \end{tabular}

\end{table*}\label{tab:circvel_atSun}

Assuming $V_{\odot, LSR} = $12.24$ \,\rm km/s$ as measured in~\cite{10.1111/j.1365-2966.2010.16253.x} from the Hipparcos Catalogue, we obtain $V_{LSR}=7 \pm 7$ km/s. Our estimate is still statistically compatible with zero streaming motion, but nevertheless strengthens the hypothesis that a region around the Sun, with a characteristic length scale of 1 kpc, exhibits a bulk motion in the azimuthal direction of the order of 10 km/s. 

\subsection{Cautionary tale about distances} \label{results_distances}

Our study is based on GSP-Phot distances~\citep{2022arXiv220606138A}, which have been shown to systematically underestimate the distance beyond 2 kpc from the Sun \citep{2022arXiv220605992F}. This bias is also shown in the left panel of Fig.~\ref{fig:heliocentric_distances}, as the two-dimensional distribution of the estimated distance versus inverse parallax is not symmetric with respect to the 1:1 line, as expected from a Gaussian noise model for parallax measurements, but is more populated to the right of this line. \cite{2022arXiv220605992F} find that imposing a cut-off in the quality of the parallax measurement of $\varpi/\sigma_{\varpi}>10$ yields reliable heliocentric distances up to 10 kpc. And~\cite{2022arXiv220606138A} point out that a strict parallax quality cut-off of $\varpi/\sigma_{\varpi}>20$ provides reliable distances. For our sample of RGB stars, the first cut-off eliminates the systematic underestimation of the GSP-Phot distances, although a slight overestimation of distances appears (see left panel of Fig.~\ref{fig:heliocentric_distances}). For this reason, as our fiducial run, we adopted the last strict cut, which alleviates the mild overestimation. However, in this section we describe how our results would change if we had adopted the less stringent cut-off in parallax quality or rather used photogeo distances from~\cite{2021AJ....161..147BMr} (hereinafter referred to as BJ distances). 

\begin{figure}[h]
        \centering\includegraphics[width=9.1cm]{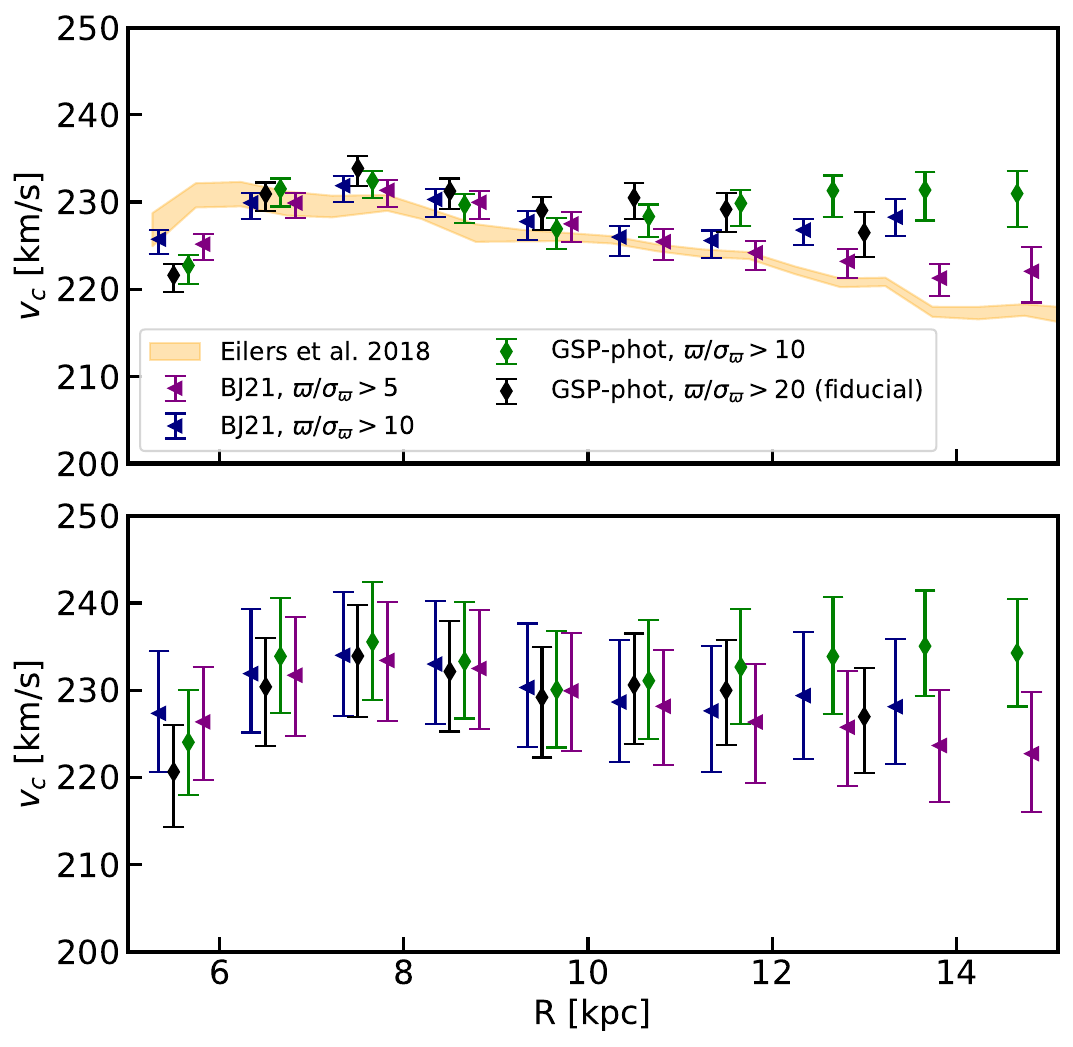}
        \caption{Circular velocity curve for different distances estimates as explained in the main text. In the top panel, we show the results where fixing the Sun's galactocentric distance to $R_0=8.122\,\rm kpc$ and leaving free the scale lengths of the radial spatial density and velocity dispersion of the tracer RGB sample. The orange band shows encompass the estimated circular velocity curve in~\cite{2019ApJ...871..120E} within 1$\sigma$. The results when additionally leaving $R_0$ as a free parameter are shown in the bottom panel. 
        \label{fig:curve_distances}
    }
\end{figure}

A bias in distance estimates has a noticeable impact in the results of our analysis. In particular, underestimated distances lead to an overestimation of the circular velocity curve gradient, and thus to an underestimation of the DM content, and vice versa in the case of overestimated distances. We note that the bias in the estimated slope is more pronounced the greater the actual slope. In order to assess the effect of biased distances, we performed our MCMC analysis using four different distance estimates: BJ distances with a cut-off of $\varpi/\sigma_{\varpi}>5$ and $\varpi/\sigma_{\varpi}>10$, and GSP-Phot distances with $\varpi/\sigma_{\varpi}>10$ and $\varpi/\sigma_{\varpi}>20$. For each of these distance estimates, the top panel of figure~\ref{fig:curve_distances} shows the resultant circular velocity curve while fixing the Sun's galactocentric distance and total velocity in the azimuthal direction to the values adopted in~\cite{2019ApJ...871..120E}, namely $R_0=8.122\,\rm kpc$ and $V_\odot=245.8\,\rm km/s$, and leaving $h_R$ and $h_\sigma$ free. The bottom panel of the same figure depicts circular velocities while letting $R_0$ as an additional free parameter. From this figure, it is clear that the inclusion of uncertainties in $R_0$ makes the four circular velocities compatible within uncertainties. Furthermore, as we increase the cut-off in the parallax quality from 5 to 10 for BJ distances, the declining of the curve becomes less steep, thus increasing the DM mass of the Galaxy. On the other hand, by increasing the cut-off from 10 to 20 for the GSP-Phot distances, we are alleviating the mild overestimation of distances and the positive gradient of the curves becomes shallower, reducing the DM mass. Table~\ref{tab:results_distances} summarises these results.

\renewcommand{\arraystretch}{1.4}
\begin{table*}
    \centering
    \caption{Summary of the results when having $R_0$, $h_R$ and $h_\sigma$ as free parameters and adopting different distance estimates. Namely, photogeo distances from \cite{2021AJ....161..147BMr} with a cut in parallax of $\varpi/\sigma_{\varpi}>5$ and $\varpi/\sigma_{\varpi}>10$, and GSP-Phot distances with quality parallax cuts of $\varpi/\sigma_{\varpi}>10$ and $\varpi/\sigma_{\varpi}>20$ (fiducial case).
    The second column shows the estimated slope of the circular velocity curve when a straight line is fitted to all bins (and in brackets the value obtained by eliminating the first two radial bins as the circular velocities increase within $\sim5-7$ kpc), the local DM density and DM mass within 14 kpc are shown in the third and forth columns, respectively. We quote the velocity of the LSR in the last column.}
    \begin{tabular}{l c c c c }
    \hline\hline
    Distance Estimate & Slope (first 2 bins removed) $[\rm km/s/kpc]$ &  $\rho_0\,\rm[GeV/cm^3]$ & $M_{DM}(R<14\,{\rm kpc})\,[M_\odot]$ & $V_{\rm LSR}\,\rm [km/s]$\\
    \hline
     BJ21 + $\varpi/\sigma_{\varpi}>5$ & $-0.9\pm0.3$ ($-1.58\pm 0.08$) & $0.37^{+0.08}_{-0.07}$  & $9.9^{+1.6}_{-1.9}$ & $7 \pm 7$\\
     BJ21 + $\varpi/\sigma_{\varpi}>10$ & $-0.3\pm0.3$ ($-1.0 \pm 0.3$) & $0.39^{+0.09}_{-0.08}$    & $10.8^{+2.0}_{-1.7}$ & $6 \pm 7$\\
     GSP-Phot + $\varpi/\sigma_{\varpi}>10$ & $0.6\pm0.3$ ($0.2 \pm 0.3$)   &  $0.43^{+0.07}_{-0.06}$ & $11.6^{+1.8}_{-2.0}$ & $6 \pm 7$ \\
     GSP-Phot + $\varpi/\sigma_{\varpi}>20$ & $0.4 \pm 0.6$ ($-1.1 \pm 0.3$) & $0.41^{+0.10}_{-0.09}$ & $11.2^{+2.0}_{-2.3}$ & $7 \pm 7$\\ 
    \hline
    \end{tabular}
\end{table*}\label{tab:results_distances}

\section{Summary and conclusions}
\label{sec:conclusions}

We estimated the circular velocity curve from 5 kpc to 14 kpc from the Galactic centre using $665\,660$ RGB stars that are approximately located in one quarter of the stellar disc with 6D phase-space information as measured by Gaia DR3, and GSP-Phot distance estimates. We determined the circular velocity curve by describing observed rotational velocities, in adimensional radial bins, as the difference between the circular velocity and the asymmetric drift. The latter given by the stationary and axisymmetric radial Jeans equations, under the further assumption of reflection symmetry above and below the Galactic plane. 
In the traditional approach, one simply plugs values into the Jeans equation. In our approach, by describing the observed rotational velocity as the circular velocity minus the asymmetric drift, we transformed the problem into an inference procedure. In particular, observed and model rotational velocities were fitted using a Bayesian inference approach that incorporates systematic and statistical uncertainties as nuisance parameters. This allowed us to propagate into the final results uncertainties of different nature. In particular, our relative uncertainties are $\sim 3\%$ and, apart from the statistics, account for uncertainties in the Sun's galactocentric distance (which is the main source of uncertainty) and uncertainties in the spatial-kinematic morphology of the stellar disc.

We studied the effect of biased distances on our results and showed, as expected, that underestimated distances lead to steeper (negative) slopes and thus to an underestimation of the dark matter content in the Galaxy. This may explain some recent findings of significantly declining circular velocity curves and, consequently, lower spherically averaged local DM densities than those purely local values obtained using stars in the Solar neighbourhood. Owing to the spherical shape of the DM halo in the inner $\sim$15 kpc of the Galaxy, these two sets of estimates should converge. 

\begin{acknowledgements} We thank the referee for her/his constructive comments and for pointing out the biases in the distance estimates. This has undoubtedly opened up many avenues for further studies.
The authors would like to thank A. Cuoco, G. Battaglia and E. Fernández Alvar for fuitful discussions and comments. 
This work was supported by the Estonian Research Council grants PRG1006, PSG700, PRG803, PSG864, MOBTP187, PRG780, MOBTT86 and by the European Regional Development Fund through the CoE program grant TK133.
This research has made use of data from the European Space Agency (ESA) mission {\it Gaia} (\url{https://www.cosmos.esa.int/gaia}), processed by the {\it Gaia} Data Processing and Analysis Consortium (DPAC, \url{https://www.cosmos.esa.int/web/gaia/dpac/consortium}). Funding for the DPAC has been provided by national institutions, in particular the institutions participating in the {\it Gaia} Multilateral Agreement.
The authors gratefully acknowledge the support of Nvidia, whose technology played a critical role in the success of this research.
GFT acknowledge support from the Agencia Estatal de Investigación del Ministerio de Ciencia en Innovación (AEI-MICIN) under grant number CEX2019-000920-S and the AEI-MICIN under grant number PID2020-118778GB-I00/10.13039/501100011033

\end{acknowledgements}

\bibliographystyle{aa}
\bibliography{references}

\begin{appendix}
\section{MCMC results} \label{App:fitting}

\begin{figure*}[h]
        \centering\includegraphics[width=17cm]{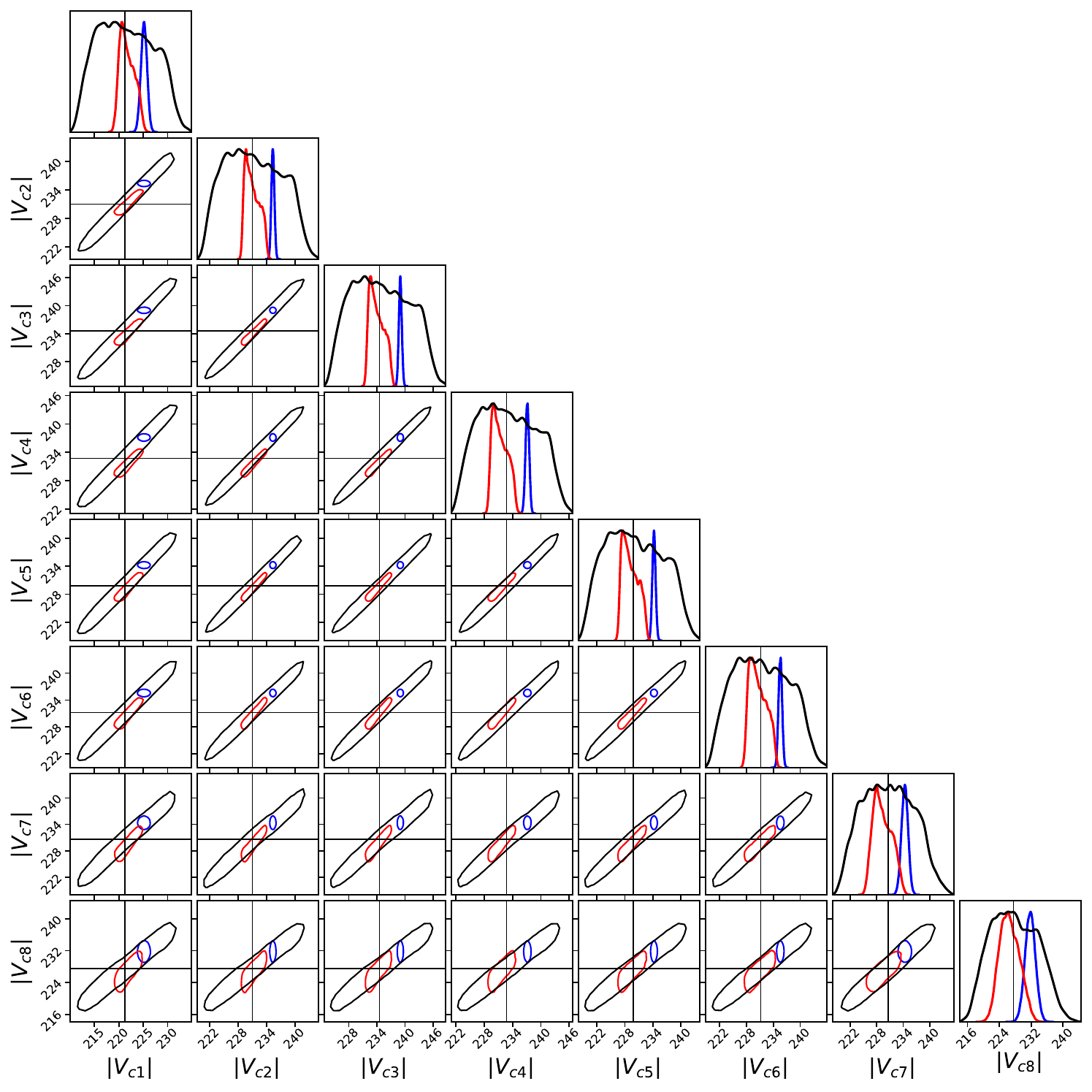}
        \caption{One and two-dimensional marginalised posterior distributions of the circular velocity. The figure shows distributions from three different parameter setups: all nuissance parameters free (black), only scale parameters free (red), all nuissance parameters fixed (blue), see text for more details. The contours delimit regions of 2-$\sigma$ probability and the best fit circular velocity of the black posteriors is demarked with horisontal and vertical lines. The diagonal contains the normalised 1D posteriors of the circular velocities from the three different runs.  
        \label{fig:corner_all_runs}
    }
\end{figure*}

Figure~\ref{fig:corner_all_runs} shows the marginalised two-dimensional and one-dimensional posterior distributions for three different MCMC runs: first, $(R_0, h_R, h_\sigma)$ are free nuisance parameters (black), second, ($h_R, h_\sigma)$ are not fixed and $R_0$ is fixed to the value $R_0=8.122\,\rm kpc$ (red) and finally, the output of the MCMC when all nuisance parameters are fixed to the values  $R_0=8.277\,\rm kpc$, $h_R=3 \,\rm kpc$, $h_\sigma=21 \,\rm kpc$.

For the first run (i.e. $R_0$, $h_R$ and $h_\sigma$ are free parameters), the movement of circular velocities is driven by changes in $R_0$. Since we are not considering strong priors on $R_0$, a strong positive correlation between this parameter and the circular velocities is observed. The circular velocity curve is sensitive to the Sun's Galactocentric distance $R_0$ (\citealt{2019JCAP...03..033B}), nonetheless, $R_0$ is constrained in the literature much better by different types of analysis than the circular velocities (see~\citealt{2013ARep...57..128M} for a review of techniques). Therefore, we do not aim to restrict $R_0$, but to assess the impact of its uncertain value on the circular velocity curve. For this reason, we adopt as a prior a uniform distribution that encompasses the most recent determinations of $R_0$ within 2$\sigma$ uncertainties. This is a conservative approach that does not favour any particular estimate. We would like to also emphasise that the actual value of $R_0$ may be troublesome. For example, the LMC is causing differences in reflex motion in distinct parts of the Galaxy causing the definition of centre to be vague or we do not know to what extent the different definitions of centre (e.g. local
isopotential curve, SMBH position) affect estimations via the Jeans equations.
One of the results of our analysis is that, given the scatter in the most recent determination of $R_0$, this parameter represents the main source of uncertainty in the calculation of circular velocities in radial bins.

On the contrary, circular velocities show a mild sensitivity to $h_R$ and $h_\sigma$. Neither of these scale lengths can be constrained by our analysis and are simply treated as nuisance parameters whose prior range is defined by observational determinations (see e.g. \citealt{2019ApJ...871..120E} and references therein).

\end{appendix}

\end{document}